\documentclass[apjl]{emulateapj}
\usepackage{apjfonts}

\begin{document}
\renewcommand{\thefootnote}{\fnsymbol{footnote}}
\title{The morphology of M17-UC1 --\\
       A disk candidate surrounding a hyper-compact \ion{H}{2} region\footnotemark[1]}
\shorttitle{The morphology of M17-UC1}

\footnotetext[1]{Based on observations made with ESO telescopes at the
La Silla  and Paranal observatories under programme IDs 71.C-0185(A),
71.C-0353(A), 73.C-0170(A), 73.C-0407(A), 75.C-0418(A), 77.C-0174(A)}

\author{M. Nielbock}
\author{R. Chini}
\author{V. H. Hoffmeister}
\author{C. M. Scheyda}
\affil{Astronomisches Institut, Ruhr--Universit\"at Bochum,
       Universit\"atsstra{\ss}e 150, 44780 Bochum, Germany}
\email{[nielbock,chini,vhoff,scheyda]@astro.rub.de}
\author{J. Steinacker}
\affil{Max--Planck--Institut f\"ur Astronomie, K\"onigstuhl 17,
       69117 Heidelberg, Germany\\
       Astronomisches Rechen-Institut am Zentrum f\"ur Astronomie
       Heidelberg, M\"onchhofstra{\ss}e 12-14, 69120 Heidelberg,
       Germany}
\email{stein@mpia-hd.mpg.de}
\author{D. N\"urnberger}
\affil{European Southern Observatory, Alonso de Cordova 3107, Vitacura,
       Santiago, Chile}
\author{R. Siebenmorgen}
\affil{European Southern Observatory, Karl-Schwarzschild-Stra{\ss}e 2,
       85748 Garching, Germany}
\email{[dnuernbe,rsiebenm]@eso.org}
\renewcommand{\thefootnote}{\arabic{footnote}}
\begin{abstract}
We investigate the morphology and the evolutionary stage of the
hyper-compact \ion{H}{2} region M17-UC1 using observations at infrared
wavelengths and NIR radiative transfer modelling. It is for the
first time resolved into two emission areas separated by a dark lane
reminiscent of an obscuring silhouette caused by a circumstellar disk. The
observational data as well as model calculations so far suggest that
M17-UC1 is surrounded by a disk of cool dust. This direct detection of a
circumstellar disk candidate around a hyper-compact \ion{H}{2} region is in
agreement with the expectations of the disk accretion model for high-mass
star formation.
\end{abstract}

\section{Introduction}
\subsection{General motivation}
The hypothesis that high-mass stars form by accretion is supported by a
number of indirect indicators like rotating molecular envelopes combined
with bipolar outflows, gas infall or maser emission towards candidate
high-mass protostellar objects (HMPOs)
\citep[e.~g.][]{ho86,keto87,keto88,cesaroni97,cesaroni05,zhang97,zhang98,
beltran04,beltran05,keto06a}. Although it seems plausible that such
rotating and collapsing configurations will eventually form a disk, the
involved observations usually lack the spatial resolution to firmly verify
such a claim. Presently, there is no convincing example of a direct detection
of a circumstellar disk around a HMPO that can be studied to infer the
details of the accretion process including the ionisation effects. Even the
currently most promising candidate, \object{IRAS 20126+4104}, must be questioned: \citet{sridharan05} report a
bipolar structure, similar to what we present in this letter. However,
\citet{cesaroni05} recently revised its mass to a value that lies just below
what is commonly accepted as ``high mass''. Furthermore, \citet{hofner99}
explained their detection of radio continuum emission as shocked gas caused
by a protostellar jet, although they also discussed free-free emission from
ionised gas as a possible source.

\begin{figure*}
\centering
\includegraphics*[width=0.80\textwidth]{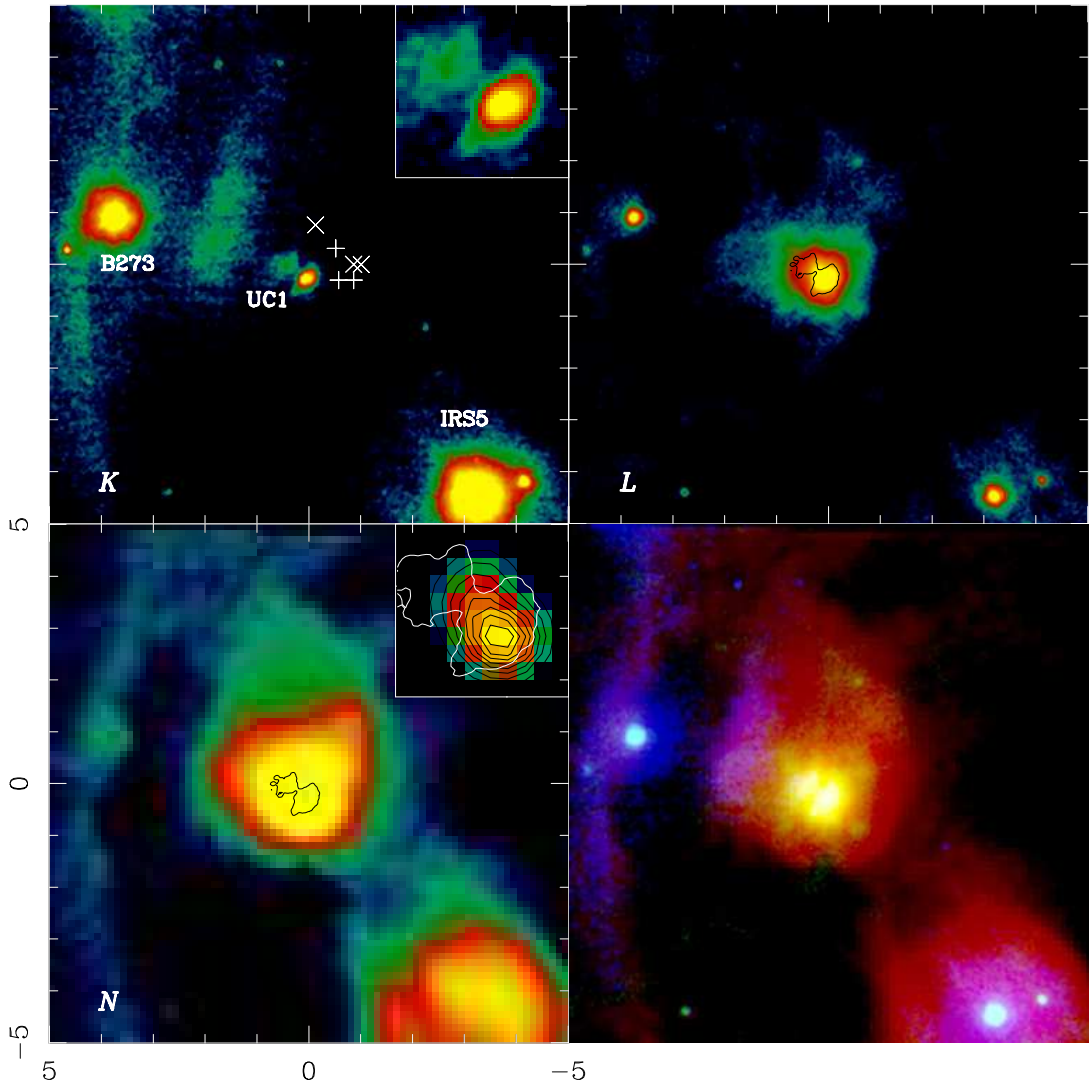}
\caption{M17-UC1 as seen by NACO ($K\!s$ and $L'$) and VISIR (SiC/$N$).
The central position is 18:20:24.82, $-$16:11:34.9 (J2000), the field
size is $10\arcsec \times 10\arcsec$. The $3\sigma$ contour of the $K\!s$
band flux is superimposed on all images. The locations of associated OH
($\times$) and class II methanol masers ($+$) are indicated in the $K\!s$ band
image. An enlargement ($1\farcs1\times1\farcs1$) of the source is shown in the
upper right corner. The inset in the VISIR image illustrates the intensity
distribution
across the bright inner emission. Black contours at arbitrary intensity
values are shown for enhancement. The lower right panel is a RGB coded
composite of the three images (blue: $K\!s$, green: $L'$, red: $N$).}
\label{KLN-RGB}
\end{figure*}

In the present study, we investigated the
hyper-compact \ion{H}{2} region (HCHII) M17-UC1 (see Sect.~\ref{s:m17uc1}
for details). HCHII are commonly associated with the earliest stages of
high-mass star formation \citep{kurtz00,depree05}. They can be explained as
a phase in early stellar evolution, when the high-mass star begins to
ionise its molecular accretion flow. In such a stage, ionisation and
accretion can co-exist \citep{keto03,keto06b}.
We used near-IR observations in order to i) overcome the
strong extinction towards the source, and ii) utilise adaptive optics
techniques allowing for high spatial resolution. As a result, we find a
bipolar structure reminiscent of a silhouette disk against a bright
background as commonly found for low-mass protostars. Starting from this
hypothesis, we perform radiative transfer modelling to verify this claim.

\subsection{The object: M17-UC1}
\label{s:m17uc1}
\object{M17-UC1} was initially discovered as
a cometary ultra-compact \ion{H}{2} region by \citet{felli80,felli84},
who suggested a B0 - B0.5 ZAMS star
as the ionising source possessing a modelled
Lyman continuum photon flux of $\simeq 2 \times 10^{47}$\,s$^{-1}$.
It was later re-classified as a HCHII with broad
($\ge 35\,$km\,s$^{-1}$) radio recombination lines and a rising
spectral index of $+1$ between 1.4 and 43\,GHz \citep{johnson98,sewilo04}.
A number of class II methanol masers \citep{menten91b} and hydroxyl masers
\citep[e.g.][]{churchwell90,caswell97,walsh98,forster99}
are located in the vicinity of M17-UC1 being typical for the surroundings
of high-mass protostars.

Within the positional accuracy of 5\farcs7, \citet{harper76}
detected a strong compact 10.6\,$\mu$m source (IRc2) coinciding with
M17-UC1 with an SED suggesting the presence of hot dust. The IR flux
was assumed to originate
entirely from M17-UC1 \citep{felli84}. When \citet{felli87} observed
M17-UC1 from 1.25 to 18.1\,$\mu$m, they also found an unresolved IR
source which they attributed to the HCHII. However, they used the strongest
$K$~band peak as their positional reference, which in fact is not M17-UC1
but the nearby IR-bright star IRS\,5 \citep{chini98}. It is dominating the
emission at $JHK$. Additional IR imaging was carried out by
\citet{giard94}, \citet{chini00} and \citet{jiang02}.
\citet{kassis02} 
characterised the source as a ZAMS B0 type surrounded by a
shell of $0.6-3.4\,M_\odot$ and a radius of less than 6600\,AU.

\section{Observations and Calibrations}
The $JHK\!sL'$ adaptive optics imaging was carried out in June 2003
using \mbox{NAOS/CONICA} \citep{naco1,naco2} at the ESO VLT. The pixel
resolution was 0\farcs027, the limiting magnitudes are $J=20.6$, $H=19.3$,
$K\!s=18.4$ and $L'=16.2$. The photometric calibration was obtained
from our ISAAC observations in September 2002 \citep{chini04b,hoffmeister06}.
We adjusted the astrometry in the $JHK\!sL'$ images
by referencing the NACO sources with the detections of the ISAAC data that
were astrometrically calibrated using the 2MASS database.
With this procedure, we estimate a relative astrometric
accuracy of better than $\pm0\farcs1$.

\setcounter{footnote}{1}
The TIMMI2 \citep{timmi2} NIR/MIR imaging was carried out during 3 observing
runs in July 2003, April 2004, and July 2005 at the ESO 3.6\,m telescope
at La Silla, Chile. The observations covered the $L$, $M$, $N1$,
$N10.4$, $N11.9$ and $Q1$ bands. A standard chopping and nodding technique
was used with an amplitude of 10\arcsec. All data are limited by diffraction
with a FWHM of 0\farcs7.

M17-UC1 was imaged with VISIR \citep{visir4} in May 2006 through the SiC
filter. The pixel scale was 0\farcs127 resulting in a
field-of-view of $32\farcs5 \times 32\farcs5$. A standard chopping
and nodding technique in perpendicular directions with throw
amplitudes of 15{\arcsec} was applied. The measured image quality
(FWHM) of $\simeq 0.32\arcsec$ is limited by diffraction.
The astrometric calibration was achieved by fitting transformation
equations between the NACO $L'$ data and the VISIR data after identifying
several point-sources visible in both wavebands. Image restoration of the
TIMMI2 and VISIR data was achieved with the software MOPSI (maintained by
R. W. Zylka, IRAM, Grenoble, France). The overall astrometric
accuracy is demonstrated by the RGB coded $KLN$ composite image presented in
Fig.~\ref{KLN-RGB}.

The $K$ band spectroscopy was obtained in August 2004 with ISAAC
\citep{isaac} at a spectral resolution of 1500 and a slit width of
0\farcs3. The spectroscopy of the $N$ band silicate absorption feature
was performed with TIMMI\,2 at the ESO 3.6\,m telescope at La Silla, Chile in
July 2003. The seeing was 0\farcs7; the slit width was 1\farcs2.

\section{Results}
\subsection{Morphology}

Fig.~\ref{KLN-RGB} displays the morphology of M17-UC1 at NIR and MIR
wavelengths. 
On our images, M17-UC1 is barely visible at $H$ with a brightness of
$17.83\pm0.25$\,mag. Surpassing previous NIR studies, we resolve the
source for the first time into two $K$ band emission blobs, separated
by $0\farcs46$ at a position angle of $126^\circ$ measured clockwise
from north to south. A dark lane separates the two $K$
band nebulae (Fig.~\ref{KLN-RGB}). The south-western emission has
an elliptical shape of $0\farcs9 \times 0\farcs5$ ($3\sigma$
contour) and a pronounced peak at 18:20:24.83, $-$16:11:35.0 (J2000)
with a FWHM (full width at half maximum) of $0\farcs19\times0\farcs13$.
The north-eastern emission at
18:20:24.85, $-$16:11:34.7 (J2000) is more diffuse and has a size of
$0\farcs8 \times 0\farcs5$. The integrated brightness of both blobs
within a radius of 3600\,AU is $K\!s = 13.1$\,mag, their intensity
ratio is 10:1 (SW:NE). At $L$ band, the source attains a spherical shape
with a radius of $\sim 1\farcs0$ ($3\sigma$ contour) and an $L'$
brightness of 6.1\,mag within 3600\,AU. 

On previous MIR images \citep{chini00,kassis02}, M17-UC1 appeared as
a point-like source with spherically symmetric circumstellar
emission. Our new $N$ band data resolve M17-UC1 (see inset in
Fig.~\ref{KLN-RGB}). The peak brightness is centred on the south-western $K$
band peak, and shows an elongation with a FWHM of $0\farcs7 \times 0\farcs5$
towards northeast being compatible with with north-eastern $K$ band peak.
We derive a flux density
of $29.87 \pm 0.13$\,Jy for the central compact source. The total
circumstellar emission ($\sim 5\arcsec\times 3\arcsec$) is relatively complex
with a noticeable extent to the northwest. It adds another $3.52$\,Jy to
the total flux density.

\subsection{Spectral appearance}

Our $K$ band spectrum (not shown) of M17-UC1
includes both nebulae and shows an extremely red continuum caused by hot dust.
Apart from the Br$\gamma$ line emission, which most likely originates
from scattered photons emitted by the gas inside the HCHII, the spectrum
appears featureless.

\begin{figure}
\plotone{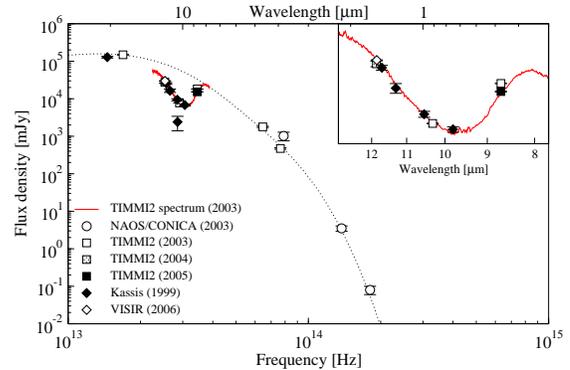}
\caption{SED of M17-UC1. The data were obtained between 1999
and 2006. The red solid line is the MIR spectrum of the silicate absorption
feature at $9.7\,\mu$m. The inset is a zoom into the spectrum.
A continuum fit of two black-body components with formal colour
temperatures of $T=425$\,K and $225$\,K is shown as a dotted line.
}
\label{SED}
\end{figure}

In the $N$ band, M17-UC1 displays a deep silicate feature
(Fig.~\ref{SED}); its existence was already suggested by earlier
photometric data \citep{harper76,kassis02}. This absorption
indicates the presence of cool dust along the line of sight towards
M17-UC1. Using the standard relation for converting optical depths of
$\tau_V \sim 17 \cdot \tau_{9.7}$ \citep{kruegel03}, we infer a visual
extinction of about 40\,mag. The luminosity between
1.6 and 20\,$\mu$m corrected for extinction is
$1.1 \times 10^4\,L_\odot$ being consistent with an early B-type star.
All our broad-band photometric data from previous epochs that were
obtained with different filters trace the shape of the silicate absorption
feature fairly well. This rules out possible MIR flux variations that were
suggested by \citet{nielbock01}.
Although our broad $N$ band imaging filter also includes the silicate
feature, we only see the extended warm emitting dust in Fig.~\ref{KLN-RGB}.

\section{Discussion}

\subsection{Observational evidence for a disk}
The symmetric absorption pattern of M17-UC1 in the $K$ band along with the
perpendicularly scattered light nebulosities on both sides very much
resembles the appearance of young low-mass stars with circumstellar disks
\citep[e.g][]{mccaughrean96,padgett99,brandner00,grosso03}.
This view is supported by the morphology of theoretically derived
synthetic images of protostellar disks presented by \citet{stark06}.
These images are based on varying disk radius and accretion rate,
among which one can find suiting representations of the $K\!s$ image of
M17-UC1. In addition, the class II methanol masers that are
located in the general direction of the supposed disk support our
conjecture.

Apart from the almost uniformly extended $L$ and $N$ band emission
one would expect to see the supposed disk to be emitting strongly in the
MIR, too. Nevertheless, there must be cool dust inside the disk as witnessed
by the silicate absorption feature. We interpret
the elongated shape of the central MIR emission perpendicular to the disk
orientation as scattered light from the central source. Due to the lower
optical depth, however, the MIR disk emission is veiled by the silicate
absorption and scattered MIR photons. At larger distances from the
HCHII, we see a warmer and less dense envelope strongly emitting in the
MIR.

Interestingly, stellar MIR emission is not
necessarily attributed to disks, but can also originate from the walls
of outflow cavities where the ambient interstellar medium interacts with
the gas stream \citep{debuizer06a,debuizer07}.
In fact, the $K$ band nebulae could also be dominated by emission from an
outflow with a wide
opening angle. In this case, the narrow waist would indicate a small disk that
restricts the outflow in these directions. However, our
$K$ band spectrum does not show any hints for shocks or outflows, so we
discard this interpretation.

\subsection{Model calculations}
\begin{figure}[t]
\plotone{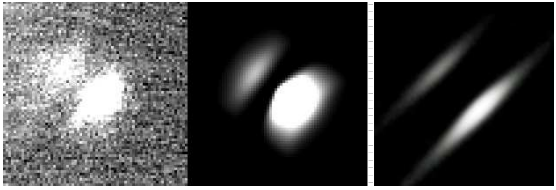}
\caption{Model fit of M17-UC1. Left: Original $K\!s$ band image.
Centre: Disk model inclined at $30^\circ$ as discussed in the text. Right: Scattered light
image at 2.2 $\mu$m from a filament in front of a B0
star. The filament has a diameter of about 3000 AU and a distance of
5000 AU to the star, the dust properties are identical to those of
the disk model.}
\label{model}
\end{figure}

We have modelled the $2.2\,\mu$m radiation with a star-disk system using a
3D radiative transfer code described by \citet{steinacker06}. The star was
assumed to have $T_{\rm eff} = 30\,000\,$K. The disk has a radial power-law
density profile with an exponential atmosphere above and below the disk
midplane like

\begin{equation}
n(r,z) = n_0 \left(\frac{r}{r_0}\right)^{\!\!\alpha} \exp\left[-\left(\frac{z}{hr}\right)^2\right]
\end{equation}

\noindent
with $r=\sqrt{x^2+z^2}$,
where $n_0$ is the number density normalisation, $h$ the scale
height, $r_0$ the inner radius, and $r_1$ the outer radius where the
disk vanishes. The $K$ band image with the triangular absorption is
well represented by parameters $n_{\rm d} = 10^4$\,m$^{-3}$,
$\alpha=1.3$, $h=0.2$, $r_0 = 20$\,AU, $r_1=1000\,$AU
(Fig.~\ref{model}) and standard $0.12\,\mu$m-sized silicate dust
particles with opacities taken from \cite{draine84}.

The inclination of the disk is about $30^\circ$ (edge-on $=
0^\circ$). We obtain a lower mass limit of $4 \times
10^{-4}\,M_\odot$ from the scattered light that only probes the
surface of the disk. A similar mass limit is derived from the
extinction that blocks the stellar light at $2.2\,\mu$m. Of course,
more mass can be hidden in the interior of the disk without
affecting the appearance of the object. Likewise, for a mildly
edge-on disk, additional dust could reside outside the scattered light
pattern.

We have also carried out radiative transfer calculations for the
alternative scenario of a dusty filament that is unrelated to M17-UC1. We
modelled dust configurations of differing sizes and
masses located at various distances in front of the star. In
contrast to the observed morphology, the scattered light generally extends
farther out along the filament, and the intensity ratio between the
strong south-western peak and the faint north-eastern peak cannot be
reproduced satisfactorily (Fig.~\ref{model}). Therefore, the radiative
transfer calculations also favour the disk model over the filament model. 

\section{Conclusion}
We have presented new NIR and MIR observations of the hyper-compact
HII region M17-UC1. As a prominent feature, the $K\!s$ band image shows
a dark lane in scattered light producing a substantial silicate absorption
feature observed in the MIR. Analysing the image with radiative
transfer models, we find the best agreement by assuming a disk-like
structure around the central source instead of a foreground filament.
For this reason and because of the "high mass" nature of the embedded star
as determined by many independent measurements, we suggest that M17-UC1
might be the first doubtless candidate of a HCHII where at least parts of
a circumstellar disk are still present.

\acknowledgments
This work was partly funded by the Nordrhein-Westf\"alische Akademie der
Wissenschaften. M.~N. acknowledges the support by the Deutsche
Forschungsgemeinschaft, project SFB 591 and the Ministerium f\"ur
Innovation, Wissenschaft, Forschung und Technik (MIWFT) des Landes
Nordrhein-Westfalen.
We thank E.~Churchwell, H.~Beuther and M.~Haas
for helpful discussions.
Likewise, we thank the referee E.~Keto for his exemplary and
valuable suggestions that significantly improved the manuscript. 
This publication makes use of data products from the Two Micron All Sky
Survey, which is a joint project of the University of Massachusetts and the
Infrared Processing and Analysis Center/California Institute of Technology,
funded by the National Aeronautics and Space Administration and the
National Science Foundation.


\end{document}